\documentclass[aps,amsfonts,amsmath,prd,preprint,nofootinbib,tightenlines,showpacs]{revtex4}
\usepackage{epsf}

\def\be{\begin{equation}}
\def\ee{\end{equation}}
\def\bea{\begin{eqnarray}}
\def\eea{\end{eqnarray}}
\def\bml{\begin{subequations}}
\def\blea{\bml\begin{eqnarray}}
\def\elea{\end{eqnarray}\end{subequations}}

\begin{document}

\title{Energy conditions for a generally coupled scalar field outside a
reflecting sphere}

\author{Delia Schwartz-Perlov}
\email{Delia.Perlov@tufts.edu}
\author{Ken D.\ Olum}
\email{kdo@cosmos.phy.tufts.edu}
\affiliation{Institute of Cosmology,
Department of Physics and Astronomy,
Tufts University, Medford, MA  02155}

\begin{abstract}
We analyze the stress-energy tensor, and the resulting energy
conditions, for a scalar field with general curvature coupling,
outside a perfectly reflecting sphere with Dirichlet boundary
conditions.  For conformal coupling we find that the null energy
condition is always obeyed, and therefore the averaged null energy
condition (ANEC) is also obeyed.  Since the ANEC is independent of
curvature coupling, we conclude that the ANEC is obeyed for scalar
fields with any curvature coupling in this situation.  We also show
explicitly how the spherical case goes over to that of a flat plate as
one approaches the sphere.
\end{abstract}


\maketitle

\section{Introduction}

 From our everyday experience we expect the energy density of any
substance to be a positive quantity.  However, quantum field theory
allows the existence of states which have negative energy densities,
as first shown theoretically by Hendrik Casimir in
1948 \cite{Casimir}.  Casimir found that the energy density between two neutral
conducting plates was negative and resulted in an attractive force
between the plates.  This attractive ``Casimir'' force has only
recently been accurately measured \cite{Lamo97}.

The reality of negative energy densities has opened the possibility
of exotic physical situations in general relativity, including
wormholes and time travel.  With no restriction on the stress-energy
tensor, one can simply write down any desired space-time geometry and
solve Einstein's equations in reverse to find the matter stress-energy
that will produce that geometry.  To prevent exotic phenomena requires
energy conditions that all matter must obey.

The simplest of these is the weak energy condition (WEC), which states
that no observer sees negative energy density,
\be\label{eqn:WEC} T_{\mu\nu}V^\mu V^\nu \geq {0}\,, \ee where
$T_{\mu\nu}$ is the stress-energy tensor and $V^\mu$ is any timelike
vector.  If instead we consider null $V^\mu$ in Eq.\ (\ref{eqn:WEC}),
we get the null energy condition (NEC).  If the WEC is obeyed at some
location for all $V$, then by continuity the NEC must also be obeyed.
These conditions, however, are violated by the Casimir effect between
parallel plates.

The averaged weak energy condition (AWEC) allows the WEC to be
violated locally, but holds that when one integrates the WEC over
a complete geodesic with tangent vector $V^\mu$, the result must
be non-negative, \be\label{eqn:AWEC} \int_{-\infty}^\infty dx\,
T_{\mu\nu}V^\mu V^\nu \geq {0}\,. \ee When $V^\mu$ is null,  Eq.\
(\ref{eqn:AWEC}) represents the averaged null energy condition
(ANEC).  The AWEC is also violated by the Casimir effect, but the ANEC
is obeyed, because the geodesic cannot pass through the plates
(considered as ideal boundaries) and so must run parallel to them, in
which case all contributions cancel.

The ANEC is the most important condition for our purposes, since its
validity is sufficient to rule out many exotic phenomena, such as
traversable wormholes \cite{Morris88b}, superluminal travel
\cite{Olum:1998mu}, and closed timelike curves \cite{cpc}, and to
prove singularity theorems
\cite{Galloway,Roman86,Roman88}\footnote{For superluminal travel, we
integrate only over the path to be traveled, and for singularity
theorems only over the part of the geodesic to the future of a trapped
surface.}. No ANEC violations are known for a scalar field in flat
space quantum field theory. In \cite{Perlov:2003}, we computed the
stress-energy tensor outside a Dirichlet sphere and found that ANEC
was always obeyed.  Here, we extend this work to arbitrary curvature
coupling.

The stress energy tensor for an arbitrarily coupled scalar field
outside a sphere was found by Saharan \cite{Saharian:2000mw}.  We
reproduce these results using a similar technique, based on the
calculational framework developed in \cite{Graham:2002yr}.  In Secs.\
\ref{sec:nec} and \ref{sec:vev} we obtain a general expression for the
vacuum stress-energy tensor outside a spherically symmetric scattering
center. In Sec.\ \ref{sec:hard} we specialize to perfectly reflecting
boundary conditions on a sphere. In Sec.\ \ref{sec:larger}, we show
how our results give analytic expressions for the stress-energy tensor
in the large $r$ limit.  In Sec.\ \ref{sec:results} we calculate the
NEC numerically and find that it is always obeyed for conformal
coupling.  Since the NEC is obeyed for conformal coupling at every
point outside the sphere, we know that the ANEC is obeyed too.
Furthermore, the ANEC is coupling constant independent, so we can
conclude that the ANEC is obeyed for all scalar field couplings.

In Sec.\ \ref{sec:smallr} we show how our results approach the usual
stress-energy tensor for a flat Dirichlet plate for points very close
to the sphere. We follow with conclusions in Sec.\
\ref{sec:conclusions}.


\section{Stress-energy tensor for general coupling}
\label{sec:nec}

The stress-energy tensor for a massless real scalar field is
\be \label{eqn:stressenergytensor}
T_{\mu\nu}=\partial_\mu\phi\partial_\nu\phi-\frac{1}{2}\eta_{\mu\nu}
\partial^\lambda\phi\partial_\lambda\phi+
\xi \left[\eta_{\mu\nu}\partial_\lambda\partial^\lambda
(\phi^2)-\partial_\mu\partial_\nu(\phi^2)\right]\,,
\ee
where $\xi$ is the curvature coupling, and we are working in the limit
where $|\phi|\ll m_{\text{Planck}}$.  For conformal
coupling in 3+1 dimensions, $\xi = 1/6$.

If we consider a static, spherically symmetric system, the stress
tensor can be written in terms of the energy density $\rho$, the
radial pressure $p_r$ and the tangential pressure $p_\perp$.  For example,
on the $x$ axis we have $T_{00} =\rho$, $T_{xx} = p_r$, $T_{yy} =
T_{zz} = p_\perp$, with other components vanishing.

 From Eq.\ (\ref{eqn:stressenergytensor}), these components are
\bea
\rho & = & \frac{1}{2}\left[\dot\phi^2+(\partial_r\phi)^2
+2(\partial_\perp\phi)^2\right]
-\xi\left[\partial_r^2(\phi^2)  +2\partial_\perp^2(\phi^2)\right]\\
p_r & = & \frac{1}{2}\left[\dot\phi^2+(\partial_r\phi)^2
-2(\partial_\perp\phi)^2\right]
+\xi\left[2\partial_\perp^2(\phi^2)-\partial_0^2(\phi^2)\right]\\
p_\perp & = &
\frac{1}{2}\left[\dot\phi^2-(\partial_r\phi)^2\right]
+\xi\left[\partial_\perp^2(\phi^2)+\partial_r^2(\phi^2)-\partial_0^2(\phi^2)\right]\,,
\eea
where $\partial_\perp$ denotes the derivative along a tangential line.
(By rotational symmetry the choice of line will not matter.)
The vacuum expectation value
$\langle\phi^2\rangle$ depends only on $r$, since the system is
static and rotationally symmetric.  Thus
$\langle\partial_0^2(\phi^2)\rangle
=\partial_0^2\langle\phi^2\rangle = 0$.  The second derivative in
the tangential direction does not vanish, however.  There is still
a term $\partial_\perp^2\langle\phi^2\rangle =
(1/r)\partial_r\langle\phi^2\rangle$.  Thus \bea
\label{eqn:rho} \rho & = &
\frac{1}{2}\left[\dot\phi^2+(\partial_r\phi)^2
+2(\partial_\perp\phi)^2\right]
-\xi\left[\partial_r^2(\phi^2)  +\frac{2}{r}\partial_r(\phi^2)\right]\\
\label{eqn:radialpressure} p_r & = &
\frac{1}{2}\left[\dot\phi^2+(\partial_r\phi)^2
-2(\partial_\perp\phi)^2\right]
+\frac{2\xi}{r}\partial_r(\phi^2)\\
\label{eqn:perppressure} p_\perp & = &
\frac{1}{2}\left[\dot\phi^2-(\partial_r\phi)^2\right]
+\xi\left[\frac{1}{r}\partial_r(\phi^2)+\partial_r^2(\phi^2)\right]
\eea

If we are interested only in the null energy condition, then the
terms proportional to the metric do not contribute, and we have
\be\label{eqn:nec1}
T_{\mu\nu}V^\mu
V^\nu=\left(V^\alpha\partial_\alpha\phi\right)^2 -\xi V ^\mu
V^\nu\partial_\mu\partial_\nu(\phi^2)\,.
\ee
If $V = (1,{\bf v})$
with $|{\bf v}|= 1$, then

\be \label{eqn:nectmunu} T_{\mu\nu}V^\mu
V^\nu=\dot\phi^2+v_r^2(\partial_r\phi)^2
+v_\perp^2(\partial_\perp\phi)^2-\xi\left[
v_r^2\partial_r^2(\phi^2)+\frac{v_\perp^2}{r}\partial_r(\phi^2)\right]
\ee where $v_\perp$ is the magnitude of the velocity in the
tangential direction, $v_\perp^2+v_r^2 = 1$.


\section{Vacuum expectation values}
\label{sec:vev}

In order to calculate the vacuum expectation value of the
stress-energy tensor, it is sufficient to compute the
expectation values of $\phi^2$, $\dot\phi^2$, $(\partial_r\phi)^2$, and
$(\partial_\perp\phi)^2$.

We can write the quantum field $\phi$ in spherical modes
\be\label{eqn:field}
\phi(r,\Omega, t)=
\sum_{\ell=0}^\infty\sum_{m=-\ell}^\ell
\int_0^\infty dk\,\frac{k}{\sqrt{\pi\omega}} \psi^\ell_k(r)^*Y_{\ell m}(\Omega)^*
e^{i\omega t}a_{k\ell m}^\dag
 + \text{c.c.}
\ee
where $\omega = k$ since the field is massless.

The wavefunctions $\psi^\ell_k(r)$ satisfy the
time-independent radial Schr\"{o}dinger equation
\be\label{eqn:Schrodinger}
\left(-\frac{d^2}{dr^2} - \frac{2}{r}\frac{d}{dr} +
\frac{\ell(\ell+1)}{r^{2}}\right) \psi^\ell_k(r) = k^2
\psi^\ell_k(r)\,.
\ee
normalized
\be
\label{eqn:norm}
\int_0^\infty k^2 r^2 \psi^\ell_k(r)^* \psi^{\ell}_{k'}(r) \, dr
= \frac{\pi}2 \delta(k-k')
\ee
so that the free wave function is just $\psi^\ell_k(r)= j_\ell(kr)$ where
$j$ is the spherical Bessel function $j_\ell(z) =\sqrt{\pi/(2z)}J_{\ell+1/2}(z)$.

 From Eq.\ (\ref{eqn:field}) we can compute the vacuum expectation
value of $\phi^2$.  There will be no dependence on the angular
position, so we can sum over the spherical harmonics using
\be
\sum_{m=-\ell}^\ell|Y_{\ell m}(\Omega)|^2=\frac{2\ell+1}{4\pi}
\ee
to get
\be\label{eqn:phi1}
\langle\phi^2\rangle
=\sum_\ell\frac{2\ell+1}{4\pi^2}\int_0^\infty dk\frac{k^2}{\omega}
\left(|\psi^\ell_k(r)|^2 - j_\ell(kr)^2 \right)
\ee
where we have subtracted the free wavefunctions.
No other counterterms are necessary, because we are considering
locations in empty space without any potential.

The vacuum expectation values for the time and radial derivatives are
straightforward,
\bea\label{eqn:dot1}
\langle\dot\phi^2\rangle &=&\sum_\ell
\frac{2\ell+1}{4\pi^2}\int_0^\infty dk\,k^2\omega
\left(|\psi^\ell_k(r)|^2 - j_\ell(kr)^2 \right)\,,\\
\label{eqn:phir1}
\langle(\partial_r\phi)^2\rangle &=&\sum_\ell
\frac{2\ell+1}{4\pi^2}\int_0^\infty dk\frac{k^2}{\omega}
\left(|\partial_r\psi^\ell_k(r)|^2 - (\partial_r j_\ell(kr))^2 \right)\,.
\eea

To compute the azimuthal term we need the derivative of the spherical
harmonics.  Without loss of generality we can work in the equatorial
plane where we can take $\partial_\perp Y_{\ell m}(\pi/2,\phi) =
(1/r)\partial_\phi Y_{\ell m}(\pi/2,\phi)
=(im/r)Y_{\ell m}(\pi/2,\phi)$.  Using the formula
\be
\sum_{m=-\ell}^\ell m^2|Y_{\ell m}(\pi/2,\phi)|^2=\frac{\ell (\ell+1)
(2\ell+1)}{8\pi}
\ee
we get
\be\label{eqn:angle1}
\langle(\partial_\perp\phi)^2\rangle =\sum_\ell
\frac{\ell (\ell+1) (2\ell+1)}{8\pi^2r^2}\int_0^\infty dk\,\frac{k^2}{\omega}
\left(|\psi^\ell_k(r)|^2 - j_\ell(kr)^2 \right)\,.
\ee

Outside any spherically symmetric potential, the wave functions
are given by
\be
\label{eqn:wavefn}
\psi^\ell_k(r)= \frac{1}{2}\left[{e^{2
i \delta_\ell}}h^{(1)}_\ell(kr) + h^{(2)}_\ell(kr)\right]
\ee
where $\delta_\ell$ is the scattering phase shift in the quantum
mechanical problem with the same potential, and the spherical Hankel
functions are $h^{(1,2)}_\ell(z) =\sqrt{\pi/(2z)}H^{(1,2)}_{\ell+1/2}(z)$.  Thus
\be
\label{eqn:diffwavefn}
|\psi^\ell_k(r)|^2-j_\ell(kr)^2= \frac{1}{4}\left[
\left(e^{2i \delta_\ell}-1\right)h^{(1)}_\nu(kr)^2 +\left(e^{-2i
\delta_\ell}-1\right)h^{(2)}_\nu(kr)^2\right]\,.
\ee

From the appendix of \cite{Perlov:2003} it is easy to show that when
we use Eq.\ (\ref{eqn:diffwavefn}) in Eqs.\
(\ref{eqn:phi1}--\ref{eqn:phir1},\ref{eqn:angle1}) we can drop the second term
of Eq.\ (\ref{eqn:diffwavefn})
and extend the range of integration over $k$ to $-\infty$, with
the understanding that $k$ is to be taken above any branch cut on
the negative real axis.

Then, following the methods used in \cite{Graham:2002yr} and
\cite{Perlov:2003} (a similar procedure was used previously in
\cite{Saharian:2000mw}), we convert each expression to a contour integral
which we close in the upper half plane.  The only contribution to the
integral comes from the branch cut along the imaginary $k$ axis.  To
the right $\omega = \sqrt{k^2} = k$, but to the left $\omega = -k$.
We then use $h^{(1)}_\ell(ix) = -(2/\pi)i^{-\ell} k_\ell(x)$, where
$k_\ell$ is the modified spherical Bessel function, $k_\ell(z)
=\sqrt{\pi/(2z)}K_{\ell+1/2}(z)$.  With $k = i\kappa$, we obtain
\bea
\langle\phi^2\rangle
&=&-\sum_\ell(-)^\ell\frac{2\ell+1}{2\pi^4}
\int_0^\infty d\kappa\,\kappa
\left(e^{2i \delta_\ell}-1\right)k_\ell(\kappa r)^2\\
\langle\dot\phi^2\rangle
&=&\sum_\ell(-)^\ell\frac{2\ell+1}{2\pi^4}
\int_0^\infty d\kappa\,\kappa^3
\left(e^{2i \delta_\ell}-1\right)k_\ell(\kappa r)^2\\
\langle(\partial_r\phi)^2\rangle &=&-\sum_\ell(-)^\ell\frac{2\ell+1}{2\pi^4}
\int_0^\infty d\kappa\,\kappa^3
\left(e^{2i \delta_\ell}-1\right)k'_\ell(\kappa r)^2\\
\langle(\partial_\perp\phi)^2\rangle &=&
-\sum_\ell(-)^\ell\frac{\ell (\ell+1) (2\ell+1)}{4\pi^4r^2}
\int_0^\infty d\kappa\,\kappa\left(e^{2i \delta_\ell}-1\right)k_\ell(\kappa r)^2\,.
\eea

\section{Perfectly Reflecting Boundary Conditions}
\label{sec:hard}

For a hard sphere of radius $a$, with Dirichlet boundary conditions,
\be
{e^{2 i \delta_\ell}} = -\frac{h^{(2)}_\ell(ka)}{h^{(1)}_\ell(ka)}
\ee
so that $\phi(a) = 0$.  Thus
\be
{e^{2 i \delta_\ell}}-1 = - \frac{2j_\ell(ka)}{h^{(1)}_\ell(ka)}
\ee
and
\be
\label{eqn:boundcond}
e^{2 i \delta_\ell(i\kappa)}-1
 =(-)^\ell\pi\frac{i_\ell (\kappa a)}{k_\ell (\kappa a)}
\ee
since $j_\ell(ix) = i^\ell i_\ell(x)$ where $i_\ell(z)
=\sqrt{\pi/(2z)}I_{\ell+1/2}(z)$.

Thus
\bea\label{eqn:phi2}
\langle\phi^2\rangle
&=&-\sum_\ell\frac{2\ell+1}{2\pi^3}
\int_0^\infty d\kappa\,\kappa
\frac{i_\ell (\kappa a)}{k_\ell (\kappa a)}k_\ell(\kappa r)^2\\
\label{eqn:dot2}
\langle\dot\phi^2\rangle
&=&\sum_\ell\frac{2\ell+1}{2\pi^3}
\int_0^\infty d\kappa\,\kappa^3
\frac{i_\ell (\kappa a)}{k_\ell (\kappa a)}k_\ell(\kappa r)^2\\
\label{eqn:phir2}
\langle(\partial_r\phi)^2\rangle &=&-\sum_\ell\frac{2\ell+1}{2\pi^3}
\int_0^\infty d\kappa\,\kappa^3
\frac{i_\ell (\kappa a)}{k_\ell (\kappa a)}k_\ell'(\kappa r)^2\\
\label{eqn:angle2}
\langle(\partial_\perp\phi)^2\rangle &=&
-\sum_\ell\frac{\ell (\ell+1) (2\ell+1)}{4\pi^3r^2}
\int_0^\infty d\kappa\,\kappa\frac{i_\ell (\kappa a)}{k_\ell (\kappa
a)}
k_\ell(\kappa r)^2\,.
\eea
in agreement with \cite{Saharian:2000mw}.

From these equations we can compute any combination of stress-energy
tensor components.  For $V$ null, from Eq.\ (\ref{eqn:nectmunu}),
\bea\label{eqn:necintegral}
T_{\mu\nu}V^\mu V^\nu =
\sum_\ell\frac{2\ell+1}{2\pi^3}\int_0^\infty d\kappa\,\kappa^3
\frac{i_\ell (\kappa a)}{k_\ell(\kappa a)}& &
\bigg\{k_\ell(\kappa r)^2
-v_r^2k'_\ell(\kappa r)^2
-\frac{\ell (\ell+1)}{2\kappa^2r^2}v_\perp^2 k_\ell(\kappa r)^2\\
& &+\xi\left[\frac{v_r^2}{\kappa^2}\partial_r^2 (k_\ell(\kappa r)^2)
+\frac{v_\perp^2}{\kappa^2 r}\partial_r (k_\ell(\kappa r)^2)\right]\bigg\}\nonumber
\eea

\section{Large distance behavior}
\label{sec:larger}

Far away from the sphere, in the large $r$ regime, the stress-energy
tensor is dominated by the $\ell=0$ partial wave contribution.  For
this case the integrals in Eqs.\ (\ref{eqn:phi2}-\ref{eqn:angle2}) can
be done analytically.  We find
\bea\label{eqn:l=0phi2} \langle\phi^2\rangle
&=&\frac{\ln(1-a/r)}{8\pi^2 r^2}\approx -\frac{a}{8\pi^2 r^3} -\frac{a^2}{16\pi^2 r^4}\\
\label{eqn:l=0dot2} \langle\dot\phi^2\rangle &=&\frac{2 r -a}{32\pi^2 (r-a)^2 r^4}\approx \frac{a}{16\pi^2 r^5}+\frac{3a^2}{32\pi^2 r^6} \\
\label{eqn:l=0phir2} \langle(\partial_r\phi)^2\rangle
&=&\frac{5-6r+4(r-a)^2 \ln(1-a/r)}{32\pi^2 (r-a)^2 r^4}\approx -\frac{5a}{16\pi^2 r^5} -\frac{9a^2}{32\pi^2 r^6}\\
\label{eqn:l=0angle2} \langle(\partial_\perp\phi)^2\rangle &=& 0\,,
\eea
where the approximate forms are those necessary to compute components
to $O(r^{-6})$.

Substituting these large $r$ approximations into Eqs.\
(\ref{eqn:rho}-\ref{eqn:perppressure}), we find the following stress-energy
components

\bea \label{eqn:rholarger} \rho & = & -\frac{a}{8\pi^2
r^5}-\frac{3a^2}{32 \pi^2 r^6}+\xi\left[\frac{3a}{4\pi^2 r^5}+\frac{3a^2}{4\pi^2 r^6} \right]\\
\label{eqn:pradlarger} p_r & = & -\frac{a}{8\pi^2
r^5}-\frac{3a^2}{32 \pi^2 r^6}+\xi\left[\frac{3a}{4\pi^2 r^5}+\frac{a^2}{2\pi^2 r^6} \right]\\
\label{eqn:pperplarger} p_\perp & = & \frac{3a}{16\pi^2
r^5}+\frac{3a^2}{16 \pi^2 r^6}-\xi\left[\frac{9a}{8\pi^2
r^5}+\frac{a^2}{\pi^2 r^6} \right]\,.
\eea
The leading terms agree with \cite{Saharian:2000mw}, with the
correction that the fraction in Eq.\ (4.17) of \cite{Saharian:2000mw}
should be $(D-1)/D$ rather than $(D-2)/(D-1)$, where $D$ is the number
of spatial dimensions \cite{Saharian:personal}.

For conformal coupling all terms of order $(1/r^5)$ cancel to give
\bea
\rho =\frac{a^2}{32\pi^2 r^6}\\
p_r=-\frac{a^2}{96\pi^2 r^6}\\
p_\perp=\frac{a^2}{48\pi^2 r^6}\,.
\eea

Using Eqs.\
(\ref{eqn:l=0phi2}-\ref{eqn:l=0angle2}) in Eq. (\ref{eqn:nectmunu}) we find a
large $r$ expression for the NEC,
\bea
 T_{\mu\nu}V^\mu V^\nu &=& \frac{a}{16\pi^2 r^5}+\frac{3a^2}{32\pi^2
r^6}-v_r^2\left( \frac{5a}{16\pi^2 r^5} +\frac{9a^2}{32\pi^2
r^6}\right)\cr
& &+\xi\left[v_r^2\left(\frac{15a}{8\pi^2 r^5}+\frac{3a^2}{2\pi^2
r^6}\right)-\frac{3a}{8\pi^2 r^5}-\frac{a^2}{4\pi^2 r^6} \right]
\eea
For conformal coupling,
\be\label{eqn:conformalneclarger}
T_{\mu\nu}V^\mu V^\nu =\frac{a^2}{32 \pi^2
r^6}\left(\frac{5}{3}-v_r^2\right)> 0
\ee
so the NEC is obeyed far from the sphere.

\section{Numerical results for the NEC with conformal coupling}
\label{sec:results}
\begin{figure}
\begin{center}
\leavevmode\epsfxsize=4in\epsfbox{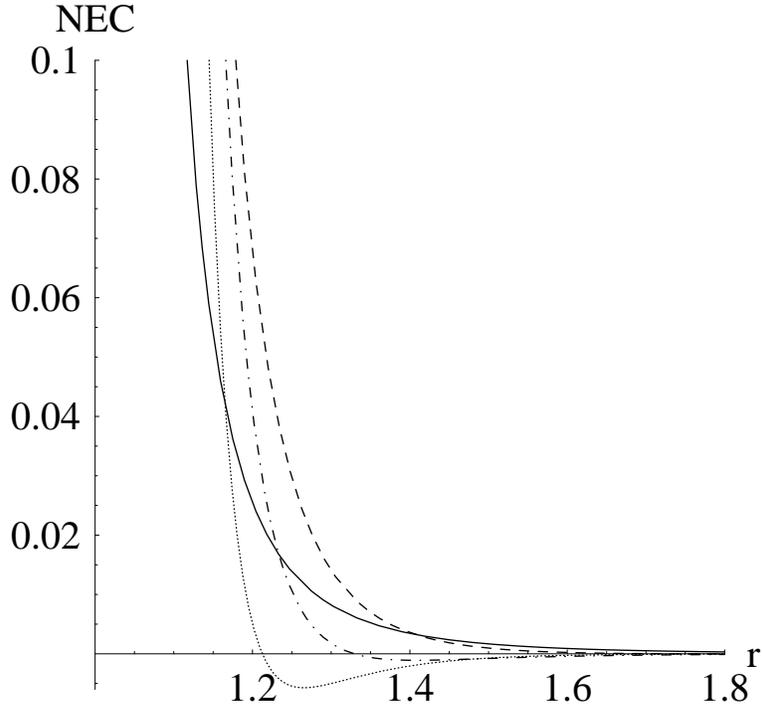}
\end{center}
\caption{We plot here the null energy condition, at a distance $r$
 from the center of a sphere for motion in the radial direction for
the $\ell=0$ (solid line), $\ell=2$ (dashed line), $\ell=4$
(dashed-dotted line), and $\ell=6$ (dotted line) partial waves.}
\label{fig:neclbyl}
\end{figure}

We have computed the NEC contribution for the case of conformal
coupling, $\xi=1/6$, by numerically performing the integral in equation Eq.\
(\ref{eqn:necintegral}).  Fig. \ref{fig:neclbyl}
shows the contributions from several partial waves for the case of
radial motion.  The case of tangential motion is very similar.  The
$\ell = 0$ contribution is always positive, as shown in Eq.\
(\ref{eqn:conformalneclarger}), but for other $\ell$, the sign depends
on position.  Thus, unlike the electromagnetic field case
\cite{Graham:2004}, the NEC does not hold in each channel
individually.

\begin{figure}
\begin{center}
\leavevmode\epsfxsize=4in\epsfbox{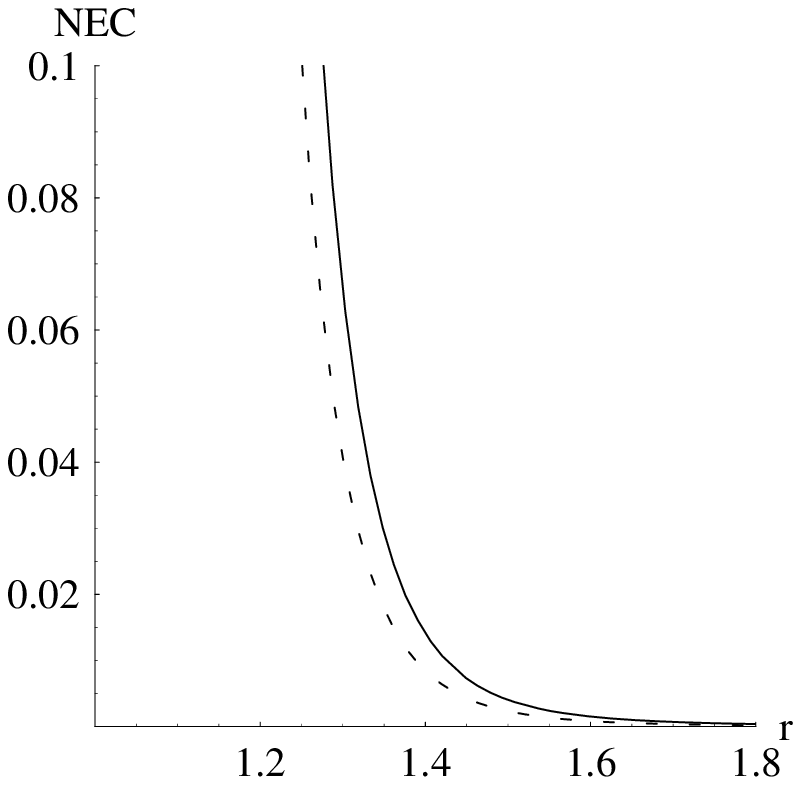}
\end{center}
\caption{The null energy condition at a distance $r$ from the
center of a sphere for motion in the azimuthal (solid line) and
radial (dashed line) directions.} \label{fig:nec}
\end{figure}

Nevertheless, when we sum up all contributions, we find that the NEC is
obeyed at any distance from the sphere, as shown in
Fig. \ref{fig:nec}.
Tangential motion yields a larger value for the NEC at any
given point than radial, but both are always positive.  Since
the NEC is obeyed at each point, the ANEC is also obeyed.

The term depending on $\xi$ in the NEC, Eq.\ (\ref{eqn:nec1}), is a
total derivative in the direction $V$, so it gives no contribution to
the ANEC integral in Eq.\ (\ref{eqn:AWEC}).  Thus the ANEC does not
depend on curvature coupling, and since it holds for conformal
coupling it will hold for any coupling, as we found numerically in the
minimally-coupled case \cite{Perlov:2003}.

\section{Perfect mirror result from spherical case}
 \label{sec:smallr}

In this section we will consider the leading-order behavior of the
stress-energy tensor when we get very close to the sphere, the
opposite limit to that of Sec.\ \ref{sec:larger}.  In this case,
there will be a divergence due to summing over large numbers of
partial wave modes, so we can use the large-order approximation
for Bessel functions, as described in \cite{Saharian:2000mw}.

The expectation value of $\phi^2$ is given by Eq.\ (\ref{eqn:phi2}),
which we can write
\be
\langle\phi^2\rangle = -\sum_\ell\frac{\nu}{2\pi^2r} \int_0^\infty
d\kappa\, \frac{I_\nu (\kappa a)}{K_\nu (\kappa a)}K_\nu(\kappa r)^2
\ee where $\nu =\ell+1/2$.

Now we will consider the case where $\epsilon = r/a-1\ll1$.  We
let $x =\kappa a/\nu$ and $y =(1+\epsilon)x =\kappa r/\nu$ to get
\be \langle\phi^2\rangle =-\sum_\ell\frac{\nu^2}{2\pi^2ra}
\int_0^\infty dx\, \frac{I_\nu (\nu x)}{K_\nu (\nu x)}K_\nu(\nu y)^2
\ee

Since we are summing over a large number of partial waves, we can
use the large-order approximation to the Bessel functions. For
$\nu\gg1$ \cite{Abramowitz}, \bea I_\nu(\nu
x)&\approx&\frac{1}{\sqrt{2\pi\nu}}
\frac{e^{\nu\eta}}{(1+x^2)^{1/4}}\\
K_\nu(\nu x)&\approx&\sqrt{\frac{\pi}{2\nu}}
\frac{e^{-\nu\eta}}{(1+x^2)^{1/4}} \eea where \be \eta(x) =
\sqrt{1+x^2} +\ln\frac{x}{1+\sqrt{1+x^2}} \ee so \be
\langle\phi^2\rangle\approx
-\frac{1}{4\pi^2a^2}\sum_\ell\nu\int_0^\infty
\frac{dx}{\sqrt{1+x^2}} e^{-2\nu(\eta(y)-\eta(x))} \ee We have
ignored higher order terms in $\epsilon$ by replacing $r$ with $a$
in the prefactor and $y$ with $x$ in the square root.

We expand to first order,
\be
\eta(y)-\eta(x)=\epsilon x\frac{d\eta}{dx} =\epsilon\sqrt{1+x^2}
\ee so we have \be \int_0^\infty \frac{dx}{\sqrt{1+x^2}}
e^{-2\nu\epsilon\sqrt{1+x^2}} =K_0(2\nu\epsilon)\,.
\ee
Thus \be \langle\phi^2\rangle \approx
-\sum_\ell\frac{\nu}{4\pi^2a^2} K_0(2\nu\epsilon)\,.
\ee
We approximate the sum by an integral,
\be\label{eqn:smallrphi}
\langle\phi^2\rangle
\approx -\int_0^\infty d\nu \frac{\nu}{4\pi^2a^2}
K_0(2\nu\epsilon) =-\frac{1}{16\pi^2a^2\epsilon^2}
=-\frac{1}{16\pi^2(r-a)^2}\,.
\ee

 From Eq.\ (\ref{eqn:dot2}), the time derivative term differs from
$\langle\phi^2\rangle$ by a factor $-\kappa^2$ in the
integrand.  By the analysis above, \be \langle\dot\phi^2\rangle\approx
\sum_\ell\frac{\nu^3}{4\pi^2a^4}\int_0^\infty dx \frac{x^2}{\sqrt{1+x^2}}
e^{-2\nu(\eta(y)-\eta(x))} \ee Using \be \int_0^\infty dx\,
\frac{x^2}{\sqrt{1+x^2}} e^{-2\nu\epsilon\sqrt{1+x^2}}
=\frac{K_1(2\nu\epsilon)}{2\nu\epsilon} \ee we find \be
\langle\dot\phi^2\rangle \approx
\sum_\ell\frac{\nu^2}{8\pi^2a^4\epsilon} K_1(2\nu\epsilon) \approx
\int_0^\infty d\nu \frac{\nu^2}{8\pi^2a^4\epsilon} K_1(2\nu\epsilon)
=\frac{1}{32\pi^2a^4\epsilon^4} =\frac{1}{32\pi^2(r-a)^4} \ee

Since $K_1(z)\sim1/z$ for small $z$, the contribution from a
single partial wave mode with $\ell\ll 1/\epsilon$ is proportional
to $\ell/\epsilon^2$.  The sum over these modes thus diverges as
$\ell^2$.  It is truncated when $\ell\sim 1/\epsilon$, so the
final result is proportional to $1/\epsilon^{4}$.

 From Eq.\ (\ref{eqn:angle2}), the tangential term differs from
$\langle\phi^2\rangle$ by a factor $\ell (\ell+1)/(2r^2)\approx
\nu^2/(2a^2)$ since we are keeping only the highest power of $\nu$
and lowest order in $\epsilon$.  Thus we find \be
\langle(\partial_\perp\phi)^2\rangle \approx -\int_0^\infty
d\nu\frac{\nu^3}{8\pi^2a^4} K_0(2\nu\epsilon)
=-\frac{1}{32\pi^2a^4\epsilon^4} =-\frac{1}{32\pi^2(r-a)^4} \ee
The angular derivative brings in a factor of $\ell$, so this term
has two more powers of $\ell$ and correspondingly two fewer powers
of $\epsilon$ in the denominator.  For $\ell\ll 1/\epsilon$ it
goes as $\ell^3\ln\nu\epsilon$, which is much smaller than the
time-derivative term, even though the sum over $\ell$ is the same
except for sign.

 From Eq.\ (\ref{eqn:phir2}), the radial term differs from
$\langle\dot\phi^2\rangle$ by a factor $-k'_\ell(\kappa r)^2/k_\ell(\kappa
r)$ in the integrand.  The derivative of the spherical Bessel function
is \be k'_\ell(z) =-\frac{\ell}{2\ell+1}k_{\ell+1}(z)
-\frac{\ell+1}{2\ell+1}k_{\ell-1}(z) \ee while the derivative of the
regular Bessel function is \be K'_\ell(z) =-\frac{1}{2}K_{\ell+1}(z)
-\frac{1}{2}K_{\ell-1}(z) \ee but the difference in the coefficients
does not matter for $\ell$ large.  Thus \be
\frac{k'_\ell(z)}{k_\ell(z)}\approx\frac{K'_\ell(z)}{K_\ell(z)}
\ee
The derivative for large order is \be K'_\nu(\nu
y)\approx-\sqrt{\frac{\pi}{2\nu}}
\frac{(1+y^2)^{1/4}}{y}e^{-\nu\eta}
=-\frac{\sqrt{1+y^2}}{y}K_\nu(\nu y) \ee Thus we find \be
\frac{k'_\ell(\kappa r)^2}{k_\ell(\kappa r)^2}\approx
1+\frac{1}{y^2} = 1+\frac{\nu^2}{\kappa^2 r^2} \ee
Consequently
\be \langle(\partial_r\phi)^2\rangle\approx
2\langle(\partial_\perp\phi)^2\rangle -
\langle\dot\phi^2\rangle \approx
-\frac{3}{32\pi^2(r-a)^4} \ee

For a conformal field we will also need \be
\partial_r^2\langle\phi^2\rangle = -\frac{3}{8\pi^2(r-a)^4}
\ee
from Eq.\ (\ref{eqn:smallrphi}),
but we will not need  $\partial_r\langle\phi^2\rangle$ because
that does not diverge as $(r-a)^{-4}$.

Putting everything together, the stress tensor components to order
$1/(r-a)^4$ are \bea \rho & \approx &
\frac{1}{2}\left[\dot\phi^2+(\partial_r\phi)^2
+2(\partial_\perp\phi)^2\right]
-\xi\partial_r^2(\phi^2)\approx -\frac{1-6\xi}{16\pi^2(r-a)^4}\\
p_r & \approx & \frac{1}{2}\left[\dot\phi^2+(\partial_r\phi)^2
-2(\partial_\perp\phi)^2\right]\approx 0\\
p_\perp & \approx &
\frac{1}{2}\left[\dot\phi^2-(\partial_r\phi)^2\right]
+\xi\partial_r^2(\phi^2)\approx\frac{1-6\xi}{16\pi^2(r-a)^4} \eea
For minimal coupling we get the standard result for a single, flat
Dirichlet plate, and for conformal coupling, $\xi = 1/6$, we also
get the standard vanishing result for the flat plate.  Therefore,
by linearity, the result is correct for general coupling.

\section{Conclusions}
\label{sec:conclusions}

We have studied the problem of a real, massless scalar field
outside a reflecting sphere, in the calculational framework
developed in \cite{Graham:2002yr}.  We obtained expressions for
the stress-energy tensor for general curvature coupling in
agreement with \cite{Saharian:2000mw}.  We analyzed the null
energy condition for the specific case of conformal coupling and
showed, analytically for large distances, and numerically
everywhere, that it is obeyed.

This implies that the averaged null energy condition is obeyed for
conformal coupling in this system.  Since the ANEC is independent of
the curvature coupling, it is obeyed for any coupling, in accord with
our conjecture in \cite{Perlov:2003} that the ANEC is satisfied for
all geodesics which pass outside any localized background potential.

We have also given a derivation of the standard stress-energy
components for a flat Dirichlet plate as the close-in limit of the
spherical geometry considered here.  The well-known divergence as one
approaches the surface arises in this case from a summation over many
angular momentum modes.

\section{Acknowledgments}

We would like to thank Noah Graham and Aram Saharian for helpful
conversations.  K. D. O. was supported in part by the National Science
Foundation.

\bibliographystyle{apsrev}
\bibliography{gr}

\end{document}